\documentclass{article}
\usepackage{graphicx}

\title{Scaling form of concentration profiles in a subdiffusive membrane system}

\author{Tadeusz Koszto{\l}owicz}
\date{\footnotesize{Institute of Physics, \'Swi\c{e}tokrzyska Academy,\\
         ul. \'Swi\c{e}tokrzyska 15, 25-406 Kielce, Poland, \\
         e-mail: tkoszt@pu.kielce.pl}}

\begin{document}

\maketitle

\begin{abstract}

We show that the concentration profiles in the subdiffusive system
with a membrane, which separates a homogeneous solution from a
pure solvent at an initial moment, has a general scaling form in
the long time limit $C\sim t^\lambda F(\delta/t^{\rho})$, where
$\delta$ is a distance from the membrane surface. There is also
derived the relation involving the subdiffusion parameters of the
medium and the membrane with the scaling parameters $\lambda$ and
$\rho$, which are measured experimentally. The relation allows one
to extract the subdiffusion parameters of the system from
experimental data.

\end{abstract}

\date{PACS numbers: 02.50.-r, 89.75.Da}

\section{Introduction}

Subdiffusive systems can be study using the scaling method, which
assumes that the concentration of a transported substance $C(x,t)$
is of the form
    \begin{equation}\label{1}
C(x,t)=t^\lambda F(z) ,
    \end{equation}
where
    \begin{equation}\label{2}
z\sim \frac{x-x_0}{t^\rho} ,
    \end{equation}
$x_0$ is an arbitrary choosen point (here we consider the
one-dimensional system). The parameters $\lambda$ and $\rho$ are
controlled by subdiffusion parameters of the system. Since the
parameters $\lambda$ and $\rho$ can be measured experimentally, it
is possible to extract the values of subdiffusion parameters of
the system from experimental data. This method was used already
for subdiffusion in homogeneous systems \cite{al,m} and in the
systems with chemical reactions \cite{yal}. Recently the
concentration of the form (\ref{1}) was observed experimentally in
the membrane subdiffusive system \cite{d1}.

We study the subdiffusive system of two vessels separated by a
thick membrane. At an initial moment one vessel contains a
homogeneous solute whereas in the second one there is a pure
solvent. Such a system has experimentally studied
\cite{d1,dwk,dkwm,d}. On the basis of theoretical model we show
that the concentrations are given by the scaling functions
(\ref{1}) and (\ref{2}). We also derive useful relations combining
the subdiffusion parameters of the membrane and solvent with the
parameters $\lambda$ and $\rho$.

\section{The method}

Let us consider the one-dimensional system with a symmetrical
membrane which is perpendicular to the $x$ axis where $x=0$ and
$x=d$ are the positions of the membrane surfaces. In the regions
$(-\infty,0)$ and $(d,\infty)$ there is a subdiffusion with the
subdiffusion parameter $\alpha$ and subdiffusion coefficient
$D_\alpha$, inside the membrane a subdiffusion occurs with the
parameters $\beta$ and $D_\beta$. We assume that the transport of
particles inside the membrane is more hindered than in the
vessels, so $\alpha>\beta$.

The concentration is described by the subdiffusion equation with
fractional Riemann-Liouville time derivative \cite{mk}
    \begin{displaymath}
\frac{\partial C(x,t)}{\partial t}
=D_{\eta}\frac{\partial^{1-\eta}}{\partial
t^{1-\eta}}\frac{\partial^{2} C(x,t)}{\partial x^{2}}\;,
    \end{displaymath}
where $\eta=\alpha$ outside the membrane and $\eta=\beta$ inside
it. To solve the subdiffusion equation we take the following
boundary conditions:
    \begin{equation}
    \label{bc1}
J(0^-,t)=J(0^+,t),
    \end{equation}
    \begin{equation}
    \label{bc2}
C(0^-,t)=\lambda C(0^+,t),
    \end{equation}
    \begin{equation}
    \label{bc3}
J(d^-,t)=J(d^+,t),
    \end{equation}
    \begin{equation}
    \label{bc4}
\lambda C(d^-,t)=C(d^+,t),
    \end{equation}
where $J$ denotes the subdiffusive flux \cite{mk}, the
dimensionless parameter $\lambda$ controls the permeability of the
membrane. We note that the boundary conditions
(\ref{bc1})-(\ref{bc4}) were already studied in the case of normal
diffusion in membrane systems \cite{hgvs}. The initial condition
we is chosen as:
\begin{displaymath}
    C(x,0)=\left\{ \begin{array}{cc}
             C_{0}, & x<0, \\
             0, & x>0.
           \end{array} \right.
\end{displaymath}

The solutions of the subdiffusion equations in terms of the
Laplace transforms are
\begin{equation}\label{f1}
\hat{C}_{1}(x,s)=\frac{C_{0}}{s}-\frac{C_{0}}{s}e^{-\sqrt{s^\alpha/D_\alpha}(-x)}
\left[\frac{1+\gamma-(1-\gamma)e^{-2\sqrt{s^{\beta}/D_{\beta}}d}}{(1+\gamma)^{2}
-(1-\gamma)^{2}e^{-2\sqrt{s^{\beta}/D_{\beta}}d}}\right] ,
\end{equation}
\begin{equation}\label{f2}
\hat{C}_{M}(x,s)=\frac{C_{0}\gamma}{\lambda s}
\left[\frac{(1+\gamma)e^{-\sqrt{s^{\beta}/D_{\beta}}(x)}+
(1-\gamma)e^{-\sqrt{s^{\beta}/D_{\beta}}(2d-x)}}{(1+\gamma)^{2}
-(1-\gamma)^{2}e^{-2\sqrt{s^{\beta}/D_{\beta}}d}}\right] ,
\end{equation}
\begin{equation}\label{f3}
\hat{C}_{2}(x,s)=\frac{2C_{0}\gamma}{s}
e^{-\sqrt{s^\alpha/D_\alpha}(x-d)-\sqrt{s^{\beta}/D_{\beta}}d}
\left[\frac{1}{(1+\gamma)^{2}
-(1-\gamma)^{2}e^{-2\sqrt{s^{\beta}/D_{\beta}}d}}\right] ,
\end{equation}
where $\gamma=\lambda
\sqrt{D_\alpha/D_\beta}s^{(\beta-\alpha)/2}$, the indexes $1$,
$M$, and $2$ are assigned to the regions $(-\infty,0)$, $(0,d)$,
and $(d,\infty)$, respectively. To obtain the inverse Laplace
transform $L^{-1}$, at first we find the series expansion of the
functions (\ref{f1})-(\ref{f3}) in terms of
$s^{\nu}e^{-as^{\beta}}$, and next we use the following formula
\cite{k1}
    \begin{equation}\label{ilt}
L^{-1}\big(s^{\nu}e^{-as^{\mu}}\big)=
\frac{1}{t^{1+\nu}}F_{\nu,\mu}\left(\frac{a}{t^\mu}\right),
    \end{equation}
where
    \begin{displaymath} F_{\nu,\mu}\left(\frac{a}{t^\mu}\right)=
-\frac{1}{\pi}\sum_{k=0}^{\infty}\frac{\sin \left [\pi\left
(k\mu+\nu \right )\right]\Gamma \left (1+k\mu +\nu\right
)}{k!}\left (-\frac{a}{t^{\mu}}\right )^{k}\;,
    \end{displaymath}
$a>0$, $\mu>0$, and the parameter $\nu$ is not limited.

Since the membrane used in experiments was not transparent for the
laser beam and consequently the concentration profiles inside the
membrane were not known), we focus in the following on the regions
outside the membrane. To simplify the calculations we consider the
above functions in the long time limit. According to the Tauberian
theorem, the long time limit corresponds to the small values of
the parameter $s$ in the Laplace transform. For the subdiffusive
system with relatively thin membrane described in the paper
\cite{k} the `long time` is evaluated to be longer than $100$
seconds. Let us note that in considered system
$\gamma\rightarrow\infty$ when $s\rightarrow 0$. The functions
(\ref{f1}) and (\ref{f3}) in the limit of small $s$ are
\begin{equation}\label{alt1}
\hat{C}_{1}(x,s)=C_{0}\left[\frac{1}{s}- \frac{D_\beta}{\lambda
\sqrt{D_\alpha}}
s^{-1-\beta+\alpha/2}e^{-\sqrt{s^{\alpha}/D_\alpha}(-x)}\right] ,
\end{equation}
\begin{equation}\label{alt3}
\hat{C}_{2}(x,s)=\frac{C_{0}D_{\beta}}{\lambda
\sqrt{D_\alpha}d}s^{-1-\beta+\alpha/2}e^{-\sqrt{s^{\alpha}/D_\alpha}(x-d)}.
\end{equation}
Applying the formula (\ref{ilt}) to eqs. (\ref{alt1}) and
(\ref{alt3}), we obtain
\begin{equation}\label{t1}
C_{1}(x,t)=C_{0}\left[1-\frac{D_\beta}{\lambda\sqrt{D_\alpha}d}t^{\beta-\alpha/2}
F_{-1-\beta+\alpha/2,\alpha/2}\left(\frac{-x}{\sqrt{D_\alpha
t^\alpha}}\right)\right] ,
\end{equation}
\begin{equation}\label{t3}
C_{2}(x,t)=\frac{C_0 D_\beta}{\lambda
\sqrt{D_\alpha}d}t^{\beta-\alpha/2}F_{-1-\beta+\alpha/2,\alpha/2}
\left(\frac{x-d}{\sqrt{D_\alpha t^\alpha}}\right).
\end{equation}

To illustrate the functions (\ref{t1}) and (\ref{t3}), there are
shown few plots of the concentration profiles for different times
in Fig.1. Here $\alpha=0.75$, $\beta=0.5$, $D_\alpha =1\cdot
10^{-3}$, $D_\beta =1\cdot 10^{-4}$, $C_0 =1$, $\lambda =2$, and
$d =1$ (all quantities are in arbitrary units).

\begin{figure}\label{fig}
\centering
\includegraphics[height=10cm]{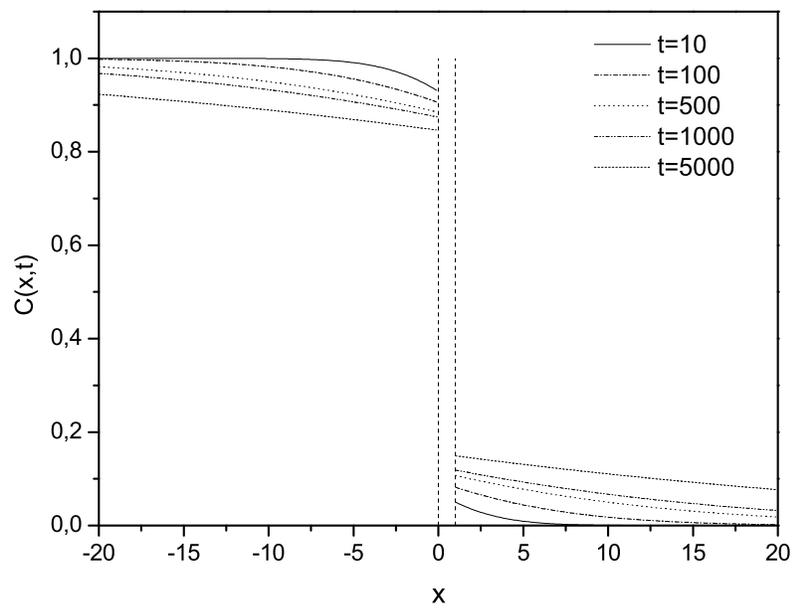}
\caption{The concentrations $C$ for different times (in
arbitrary unit). The dotted vertical lines represent the membrane
surfaces.}
\end{figure}

In the study \cite{d1} it was found that the concentration
profiles in the region $(d,\infty)$ is given by the function
(\ref{1}) with the argument (\ref{2}). The parameter $\rho$
depends on the properties of subdiffusive medium outside the
membrane and it is controlled by the subdiffusion parameter
$\alpha$ whereas the parameter $\lambda$ depends on the both of
the subdiffusive parameters $\alpha$ and $\beta$. Comparing the
function (\ref{t3}) with eqs. (\ref{1}) and (\ref{2}), we obtain
\begin{equation}\label{eq}
\beta=\lambda+\rho ,
\end{equation}
and
\begin{equation}\label{eq1}
\rho=\alpha/2 .
\end{equation}
Since the parameters $\lambda$ and $\rho$ are measured
experimentally \cite{d1}, we can get the subdiffusion parameter
inside the membrane from eq. (\ref{eq}).

\section{Final remarks}

We observe that the functions $C_0 -C_1(x,t)$ and $C_2(x,t)$ are
of the same scaling form and they can be expressed by the function
\newline $t^{\beta-\alpha/2}
F_{-1-\beta+\alpha/2,\alpha/2}\left(\frac{\delta}{\sqrt{D_\alpha
t^\alpha}}\right)$, where $\delta$ is the distance from the
membrane surface. Knowing the scaling form we derive the relations
(\ref{eq}) and (\ref{eq1}).

Presented procedure allows one to calculate the subdiffusion
parameter inside the membrane. Our theoretical result fully
coincides with the experimental one presented in \cite{d1}, where
it is reported that the concentration profiles obey the scaling
function (\ref{1}) for the system containing the membrane made of
PEG 2000 located in the agarose gel solvent, with $\rho=0.44\pm
0.01$ and $\lambda=0.17\pm 0.01$. The first parameter was
determined by means of the time evolution of near membrane layers,
the second one was extracted from the time evolution of the
concentration at the membrane surface $x=d$, which is
$C_2(d,t)\sim t^{\beta-\alpha/2}$. Substituting these values to
the relations (\ref{eq}) and (\ref{eq1}), we obtain $\beta=
0.61\pm 0.01$ for the membrane under considerations.

\section*{Acknowledgements}

The author wishes to express his thanks to Stanis{\l}aw
Mr\'owczy\'nski for fruitful discussions and critical comments on
the manuscript. This paper was supported by Polish Ministry of
Education and Science under Grant No. 1 P03B 136 30.

\end{document}